\documentstyle[epsf,float,seceq,twocolumn]{jpsj}

\relax

\author{Yoshihiro Takushima, Akihisa Koga, and Norio Kawakami}
\title{
Competing Spin-Gap Phases in a Frustrated Quantum Spin System
in Two Dimensions
}
\inst{Department of Applied Physics, Osaka University,
Suita, Osaka 565-0871}
\recdate{\today}
\abst{
We investigate quantum phase transitions among the spin-gap phases
and the magnetically ordered phases in a two-dimensional
frustrated antiferromagnetic spin system,
which interpolates several important models
such as  the  orthogonal-dimer model as well as the model on the
1/5-depleted square lattice.
By computing the ground state energy, the staggered susceptibility
and the spin gap by means of the series expansion method,
we determine the ground-state phase diagram and discuss
the role of geometrical frustration.
In particular, it is found that a RVB-type spin-gap phase
proposed recently for the orthogonal-dimer system is adiabatically
connected to the plaquette phase known for
the 1/5-depleted square-lattice model.
}
\kword{frustration, quantum phase transition, series expansion,
two dimensional spin system}
\begin{document}
\sloppy
\maketitle
\section{Introduction}
Geometrical frustration plays an important role for the
spin-gap formation in a certain class of antiferromagnetic
quantum spin systems.
Two typical spin-gap compounds found in two dimension (2D) are
the transition metal oxides, $\rm SrCu_2(BO_3)_2$\cite{Kage} and
$\rm{CaV_{4}O_{9}}$.\cite{depE1}
In the former compound, the orthogonal-dimer structure of
the $\rm{Cu^{2+}}$ spins stabilizes the dimer phase
with the spin gap, which is properly described by the 2D Heisenberg model
on the square lattice with diagonal interactions
(Shastry-Sutherland model),\cite{S-S}
as claimed by Miyahara and Ueda.\cite{Ueda1}
For the model, it was pointed out that there should exist
an intermediate phase  between the dimer
 and the antiferromagnetic phases,\cite{Mila,KogaL,2trip}
implying that strong frustration due to the competing interactions
is quite important for understanding this model.
On the other hand, the latter spin-gap compound $\rm{CaV_{4}O_{9}}$
may be described by the 2D Heisenberg model
on the 1/5-depleted square lattice\cite{depT1}
or the meta-plaquette model.\cite{Metap1,Metap2}
The spin-gap phase in this compound
results from the plaquette-singlet formation,
which is further stabilized by the frustration effect
due to the next-nearest-neighbor interactions.

An interesting feature common to the above systems is that
geometrical frustration due to the
competing interactions plays an essential role to
stabilize the disordered ground state. It is thus important
to deal with these systems in a unified framework to
clarify the role of frustration systematically.
 In particular, since a RVB-type
spin-gap phase\cite{KogaL} proposed for the Shastry-Sutherland
model may be caused by geometrical frustration,
it is desired to further clarify the nature of this spin-gap phase.
This is also important from the experimental point of view, because
the compound $\rm SrCu_2(BO_3)_2$ is expected to be located around
the phase boundary between the dimer and RVB-type phases.

Motivated by the above hot topics, we study the competition among
 the spin-gap phases as well as the magnetically ordered phases
in a 2D frustrated quantum spin system on the square lattice
with some diagonal couplings. The basic idea in this paper is the
adiabatic continuation, which allows us to relate
apparently different states which belong to the same phase
at the fixed point. By using this model, we
systematically describe quantum phase transitions
for  the orthogonal-dimer model
and the Heisenberg model on the 1/5-depleted square lattice,
where the ground-state phase diagram was already
discussed in detail.\cite{Ueda1,Mila,KogaL,2trip,KogaB,depT3}
In particular,
we show that a RVB-type spin-gap phase naturally emerges
in the Shastry-Sutherland model,
 by observing the adiabatic evolution of the spin-gap state when
the competing interactions are varied.

This paper is organized as follows.
In {\S 2}, we introduce a 2D frustrated spin model,
which allows us to treat  the above spin-gap compounds systematically.
After briefly summarizing the series expansion method in {\S 3},
we investigate the first- and second-order quantum phase transitions
among the spin-gap phases and the magnetically ordered phases
in {\S 4}. We there determine the phase diagram and clarify
the role of geometrical frustration in this class of
quantum spin systems.
A brief summary is given in {\S 5}.

\section{Model Hamiltonian}
\begin{figure}[htb]
\vspace{0.1cm}
\epsfxsize=7cm
\centerline{\epsfbox{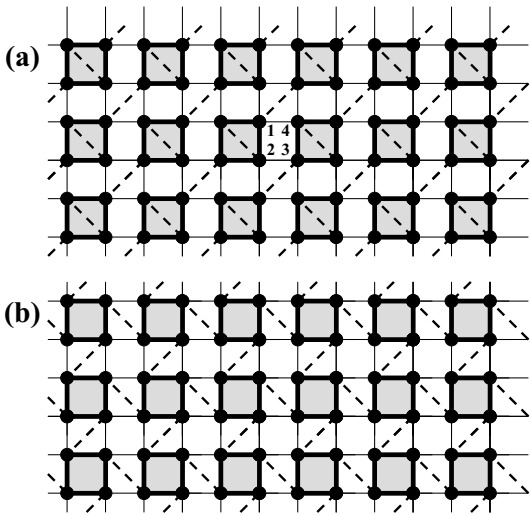}}
\caption{Frustrated spin models on the square lattice with
some diagonal bonds.
The bold, thin and  dashed lines represent the
coupling constants $J'_{1}, J'_{2}$
and $J$.
The initial configuration of the isolated dimers (plaquettes)
for the series expansion is
indicated by the dashed lines (shaded squares).}
\label{fig:md23}
\end{figure}

We consider the Heisenberg model on  the square lattice
with some diagonal couplings,
which is described by the Hamiltonian
\begin{eqnarray}
H&=&J\sum_{diagonal}{\bf S}_{i}
\cdot{\bf S}_{j}+
J'\sum_{square}{\bf S}_{i}
\cdot{\bf S}_{j},
\label{eq:model}
\end{eqnarray}
where ${\bf S}_{i}$ is the s = 1/2 spin operator at the {\it i}-th
site, and $J$, $J'_1$ and $J_2'$ represent the antiferromagnetic
exchange couplings. By putting $J$ on the orthogonal dimer
bonds (dashed lines), we then distribute the other
couplings $J'_{1}$ and $J'_{2}$
 in two ways as shown in Figs. \ref{fig:md23}(a) and \ref{fig:md23}(b).
In the case (a), there is a diagonal bond $J$ in each plaquette
formed by the coupling $J'_1$ (bold line),
while in the case (b) there is no diagonal bond in shaded plaquettes.
What is interesting in this generalization is that we can
discuss several different models in the same framework.
Namely, both models of (a) and (b) are related to each other
via the Shastry-Sutherland model with  $J'_{1}=J'_{2}$, and moreover they
 are reduced to the well-known models in the limiting case of $J'_2=0$:
the orthogonal-dimer chain for (a)
and the spin system with a windmill structure for (b). Note that
the latter is topologically equivalent to the Heisenberg model
on the 1/5-depleted square lattice for the compound $\rm CaV_4O_9$.
Therefore, by using the above models we can systematically study
 how the geometrical frustration in the 2D model affects
the generation of the spin gap.
For later convenience, we introduce the ratios $\alpha=J'_{1}/J$
and $k=J'_{2}/J'_{1}$.

\section{Series Expansion Method}

In order to study quantum phase transitions,
we make use of the series expansion method.\cite{Gel1}
This  method is powerful especially for
quantum spin systems with frustration,
where quantum Monte Carlo simulations may suffer from sign problems.
In the present context, we wish to mention that this method was successfully
applied to the meta-plaquette system,\cite{Metap1,Metap2}
the orthogonal dimer system,\cite{KogaL,2trip,KogaB,Wei} etc.
To determine the phase diagram,
we consider two initial configurations in
 the series expansion.  The first one is a
plaquette-singlet configuration composed of the shaded
squares in Figs. \ref{fig:md23}(a) and \ref{fig:md23}(b).
In this case, we parameterize
the antiferromagnetic couplings labeled by
the bold,  thin and  dashed lines as
$J'_{1}, J'_{2}=\lambda kJ'_{1}$ and $J=\lambda J'_{1}/\alpha$,
where we have introduced an auxiliary series-expansion
parameter $\lambda$ in addition to the
dimensionless couplings $k$ and $\alpha$. The original model we
are interested in
is recovered by setting $\lambda=1$.
Note that a RVB-type spin-gap phase,\cite{KogaL}
 which has the spatially-extended
disordered ground state, may be reached in this approach.
The second one is a dimer configuration, by which
we can describe the dimer phase.  A convenient
parameterization in the dimer expansion is
 $J(=1), J'_{1}=\lambda \alpha$ and $J'_{2}=\lambda k\alpha$.
In this way, these two different approaches enable
 us to perform the series  expansion
starting from isolated dimer and plaquette singlets ($\lambda=0$).

To carry out the series expansion explicitly,
 the original Hamiltonian eq.(\ref{eq:model}) is divided
into two parts,
\begin{eqnarray}
H&=&H_{0}+\lambda H_{1}\nonumber\\
&=&x\left[\sum{\bf S}_{i}\cdot{\bf S}_{j}+
\lambda\sum\Gamma_{ij}{\bf S}_{i}\cdot{\bf S}_{j}\right],\label{eq:model1}
\end{eqnarray}
where $x=J$ $(J'_{1})$ and $\Gamma_{ij}=\alpha$ or $k\alpha$ $(k$ or
$\alpha^{-1})$ for the dimer (plaquette) expansion.
The first term $H_{0}$ is the unperturbed Hamiltonian
which stabilizes  the isolated dimer or plaquette singlets, whereas
the perturbed Hamiltonian $H_{1}$
labeled by $\lambda$ connects these isolated
dimer or plaquette singlets, yielding a 2D frustrated
spin system.
We compute the staggered susceptibility $\chi_{AF}$, the spin-triplet
excitation energy $E({\bf q})$, and the ground state energy $E_{g}$ as
a power series in $\lambda$.
To estimate the susceptibility, we introduce the Zeeman term
\begin{eqnarray}
H' &=& h \left[\sum_{i\in A}{\bf S}^{z}_{i}-
\sum_{i\in B}{\bf S}^{z}_{i}\right],
\label{eq:model2}
\end{eqnarray}
where $h$ is the staggered magnetic field and $A (B)$ denotes one of
the two sublattices.

We determine the quantum phase transition point in three ways:
(i) comparison of the ground state energy, (ii)
divergence of the staggered susceptibility, (iii)
 vanishing point of the spin gap.
The latter two schemes may be applied to the second-order
phase transition, while the first one is more effective
to study the first-order phase transition.
In order to obtain the accurate phase boundary,
 we further use the Dlog Pad$\rm{\acute{e}}$
approximants\cite{Pade} for the calculated results up to the finite
order in $\lambda$.
We also make use of the {\it{biased}} Pad$\rm{\acute{e}}$
approximants,\cite{Pade}
in which the critical exponents for our 2D quantum spin
model are assumed to take those expected for the 3D classical
Heisenberg model.\cite{3Dcl}
The phase-transition point, $\lambda_{c}$, is then determined by the
formula $\chi_{AF} \sim (\lambda_{c}-\lambda)^{-\gamma}$ and
$\Delta \sim (\lambda_{c}-\lambda)^{\nu}$ with the critical exponents
$\gamma = 1.4$ and $\nu = 0.71.$\cite{Crit}

\section{Quantum Phase Transitions}
We now discuss quantum phase transitions
when the competing exchange couplings are varied.
There are three parameters in the perturbation term,
$\lambda, \alpha$ and $k$.
We first study the model given by
Fig. \ref{fig:md23}(a), which continuously connects the
2D Shastry-Sutherland model and the 1D orthogonal-dimer chain, and
then move to the model (b), which includes the
Heisenberg model on the 1/5 depleted square lattice.

\subsection{Adiabatic evolution from the orthogonal-dimer chain}

Let us start with the system shown in  Fig. \ref{fig:md23}(a).
Note that this spin system is reduced to
the orthogonal-dimer chain for $k=0$,
 which was already studied in detail:
 when the ratio $\alpha$ of the competing exchange couplings is varied,
the first-order quantum phase transition
between the spin-gap phases with the dimer and the plaquette structures
occurs at $\alpha_{c1}=0.81900$.\cite{Ivanov,KogaB}
The merit to study this system is that we can check how the
known spin-gap states in the chain system
adiabatically evolve when the system approaches
the 2D Shastry-Sutherland model.

\begin{figure}[htb]
\vspace{0.1cm}
\epsfxsize=8cm
\centerline{\epsfbox{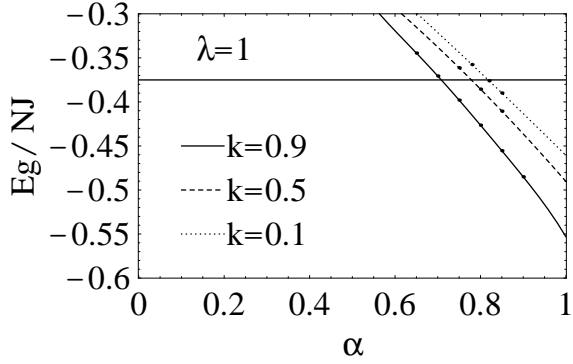}}
\caption{Ground state energy per site as a function of $\alpha = J'_{1}/J$
(for $\lambda=1$).
The flat line ($E_{g}/JN$ = -3/8) is the energy
for the exact dimer state. The dotted, dashed and  solid
lines denote the energy at $\lambda=1$ obtained
by the plaquette expansion for $k=0.1$, 0.5 and 0.9, respectively.}
\label{fig:pladene}
\end{figure}

We first discuss the effect of  interchain coupling $k$
on the first-order phase transition between two
spin-gap phases
by computing the ground state energy $E_{g}$.
Note that the energy for the dimer phase is unchanged
even on the introduction of $k$ due to the
orthogonal-dimer structure.\cite{S-S} For the plaquette phase, by
choosing the isolated plaquettes shown by shaded squares
in Fig. \ref{fig:md23}(a) as an initial configuration,
we compute the ground state energy $E_{g}$ up to the sixth order
in $\lambda$ with $\alpha$ and $k$ being fixed.
By applying the first-order inhomogeneous differential method to
the obtained series,
we then estimate the ground state energy for the plaquette phase
by setting $\lambda=1$.
The energy estimated  for $k=0.1$, 0.5 and 0.9 is
shown in Fig. \ref{fig:pladene}.
\begin{figure}[htb]
\vspace{0.1cm}
\epsfxsize=7cm
\centerline{\epsfbox{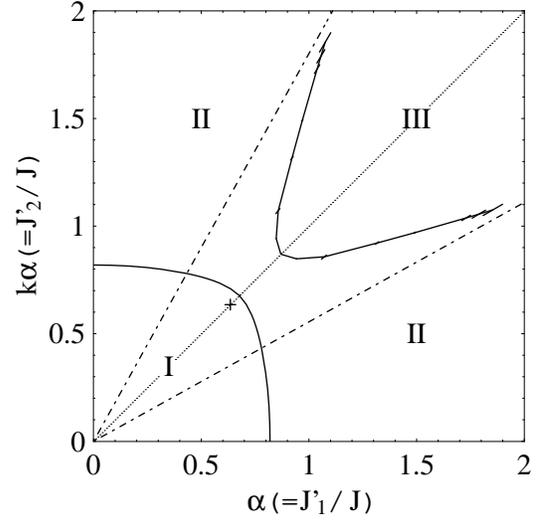}}
\caption{Phase diagram for the 2D frustrated spin system
 shown in Fig. \ref{fig:md23}(a).
The phases I, II and III represent the dimer,
 plaquette and  magnetically ordered phases, respectively.
The dotted line corresponds to the 2D Shastry-Sutherland model and
the dot-dashed lines to the asymptotic phase boundary for large $\alpha$.
The cross indicates the parameters deduced for
${\rm SrCu_{2}(BO_{3})_{2}} $.\cite{Ueda2}}
\label{fig:pladphase}
\end{figure}
It is seen that the introduction of  $k$
prefers the plaquette phase, shifting
the first-order transition point
$\alpha_{c1}$ to the left in Fig. \ref{fig:pladene}.
From these analyses,
we determine the phase boundary between the dimer phase  and
the plaquette phase in the 2D
parameter space, which is shown in Fig. \ref{fig:pladphase}.
We have checked that the present estimate of
the critical point for $k=0$ reproduces the
above-mentioned value of $\alpha_{c1}$
 fairly well. It is seen that
the phase boundary is not so sensitive to $k$ for small $k$,
 because the energy for the plaquette phase decreases
quadratically in $k$.

It is to be noted that these two competing spin-gap phases are
continuously connected to those in the Shastry-Sutherland model $(k=1)$,
with the nature of the first-order transition being
unchanged.\cite{KogaL} To check the validity of the obtained results,
we have also performed the plaquette expansion
starting from another initial
configuration, where an isolated plaquette is formed by
four spins shown as 1, 2, 3 and 4 in Fig. \ref{fig:md23}(a).
From the ground state energy computed up to the seventh order, we
confirm that the phase boundaries obtained by two distinct ways
are indeed consistent within our accuracy.

\begin{figure}[htb]
\vspace{0.1cm}
\epsfxsize=8cm
\centerline{\epsfbox{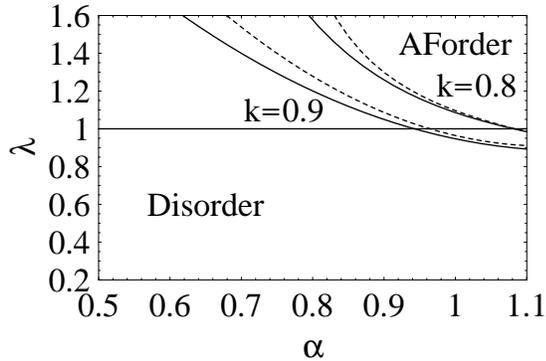}}
\caption{Second-order phase transition between the plaquette
phase and the antiferromagnetically ordered phase
for the 2D spin system given by
 Fig. \ref{fig:md23}(a) (for $k=0.9$ and 0.8).
The solid (dashed) line indicates the phase boundary obtained
by biased [3/2] ([2/2]) Pad$\rm{\acute{e}}$ approximants for
the spin gap (the staggered susceptibility).}
\label{fig:pladgs}
\end{figure}

To complete our discussions on this model,
it is necessary to discuss whether the system is driven to
the magnetically ordered phase with the increase of
the interchain coupling. To this end,
we calculate the staggered susceptibility $\chi_{AF}$ and the spin
gap $\Delta$ by means of the plaquette expansion up to the fourth
and the fifth order in $\lambda$, respectively.
We show the results  obtained for $k=0.9$ and 0.8
in Fig. \ref{fig:pladgs}.  It is seen that two lines determined
from different  quantities provide
consistent values even around  $\lambda=1$, implying
that the phase boundary may be obtained rather accurately for the
original model ($\lambda=1$) although our calculation is
restricted  to the lower-order series expansion.  From the
above analyses, we end up with  the phase
boundary shown in  Fig. \ref{fig:pladphase},
 which separates the plaquette phase II and
the magnetically ordered phase III.
Note that for large $\alpha$, the boundary between II and III
approaches the correct asymptotic line,
which was estimated as $k_c=0.56$.\cite{Neelb}
This also supports that the  present calculation
gives the reliable results for the phase diagram.

In the obtained  phase  diagram
 shown in Fig. \ref{fig:pladphase},
 we remark again that two competing spin-gap states in
the orthogonal dimer chain\cite{Ivanov,KogaB} are
continuously connected to those in
Shastry-Sutherland model.  In particular, a RVB-like
intermediate  spin-gap phase \cite{KogaL} proposed for the
Shastry-Sutherland model naturally emerges via the
adiabatic continuation of
 the plaquette phase in the orthogonal-dimer chain.

\subsection{Adiabatic evolution from the 1/5-depleted square lattice}

\begin{figure}[htb]
\vspace{0.1cm}
\epsfxsize=7cm
\centerline{\epsfbox{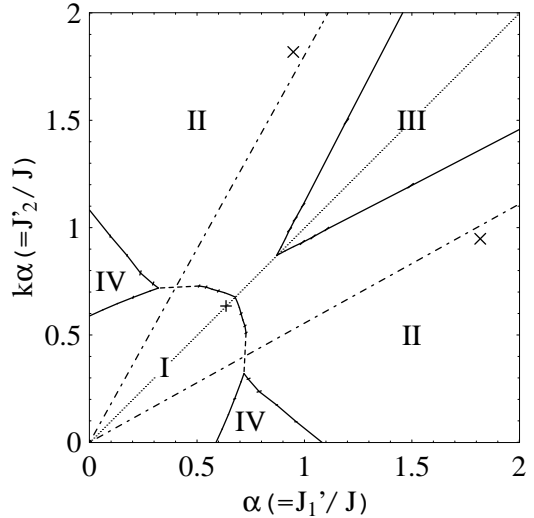}}
\caption{Phase diagram for the 2D frustrated spin system
shown in Fig. \ref{fig:md23}(b).
The phase I and II represent the dimer and the plaquette phase.
The phase III (IV) represents the antiferromagnetically ordered
phase on the square lattice (1/5-depleted square lattice).
The dotted line indicates the 2D Shastry-Sutherland model and
the dot-dashed lines the asymptotic phase boundary for large $\alpha$.
The crosses + and $\times$
indicate the location of
${\rm SrCu_{2}(BO_{3})_{2}} $\cite{Ueda2} and
${\rm CaV_{4}O_{9}}$,\cite{Metap1} respectively.}
\label{fig:depphase}
\end{figure}

We next deal with the spin system with a slightly different
structure shown in Fig. \ref{fig:md23} (b).
Recall that when $k=1$ ($k=0$), this model with $\lambda=1$
is reduced to the Shastry-Sutherland model (1/5-depleted square lattice
model).  We here study how the spin-gap phase in the
orthogonal-dimer model for  ${\rm SrCu_{2}(BO_{3})_{2}}$
is related to that in the 1/5-depleted square-lattice
model for $\rm{CaV_{4}O_{9}}$, when the
frustrating interaction $k$ is continuously varied.
We carry out the dimer (plaquette) expansion
for the ground-state energy, the staggered susceptibility
and the spin gap.  By
performing similar asymptotic  analyses mentioned  above,
the phase diagram is  determined, which
 is shown in Fig. \ref{fig:depphase}. The detail of calculation
 will be described  below for each case.

Let us start our discussions with the case of $k=0$, i.e.
 the 1/5-depleted square lattice model, which
was already studied  in connection
with $\rm{CaV_{4}O_{9}}$.\cite{depT3}
In this case, as the ratio of the exchange interactions
$\alpha$ is changed, the dimer phase I first undergoes
the second-order phase transition to the antiferromagnetically ordered
phase IV at the critical point $\alpha_{c2}^{l}=0.590(5)$,
and this ordered phase  further  shows the second-order phase
transition to the plaquette phase II at $\alpha_{c2}^{u}=1.09(1)$.
The critical points obtained by the
present method agree fairly well with those of
quantum Monte Carlo simulations.\cite{depT3}
Note that the antiferromagnetically ordered phase on
1/5-depleted square lattice labeled by IV
is different from that on the square lattice indicated by
III.

\begin{figure}[htb]
\vspace{0.1cm}
\epsfxsize=8cm
\centerline{\epsfbox{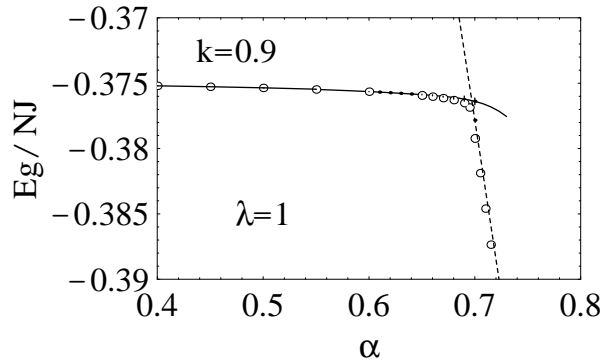}}
\caption{Ground state energy per site as a function of
$\alpha=J'_{1}/J$ (for $\lambda=1$ and $k=0.9$) for the
model shown in Fig. \ref{fig:md23}(b).
The solid (dashed) line indicates the energy
for the dimer (plaquette) phase.
We also show the results obtained by the exact diagonalization
for $N=24$ sites with a periodic boundary condition as open circles.}
\label{fig:depene09}
\end{figure}

As the coupling $k$ is increased, the spin-gap
phases become more stable while
the magnetic correlation for the phase IV is
suppressed, which is mainly due to the frustrating
nature of the interaction $k$. It is seen that the competition
among three phases  persists up to
the multicritical point $(\alpha, k\alpha)\simeq (0.72, 0.32)$
beyond which  the magnetically ordered phase IV
disappears. Thus for larger $k$
 the dimer state directly undergoes the first-order
phase transition to the plaquette spin-gap phase.
Note that the dimer phase here does not have the exact eigenstate
except for the orthogonal-dimer case $k=1$,
which is contrasted to the model (a).
We thus have  estimated the ground state energy
both of the dimer and the plaquette phases in this case.
For reference, we show in Fig. \ref{fig:depene09}
the ground-state energy calculated for $k=0.9$, where
the dimer (plaquette) expansion has been first performed up to the
eighth (seventh) order in $\lambda$,
 and then the first-order inhomogeneous differential method
has been used to obtain the values at  $\lambda=1$.
To confirm the validity of our series expansion, we have also
performed  the exact diagonalization
for the system with $N=24$ sites with a periodic boundary condition,
which is found to be  consistent with the series-expansion results.
 The phase boundary thus obtained between two spin-gap phases
is drawn in Fig. \ref{fig:depphase}.
We note that our series expansion
turns out to be not so accurate around the tricritical point, so that
the phase boundary around there is shown as the dashed line.

When the coupling  $k$ is further increased, the
plaquette phase
 shows the instability to an antiferromagnetically
ordered phase III on the square lattice.  To study this instability
from II to III, we have performed the plaquette expansion for
the staggered susceptibility (spin gap)
up to the fourth (fifth) order, and determined the phase boundary
using the biased Pad\'e approximants, which is
 shown in Fig. \ref{fig:depphase}.

Consequently, we arrive at the phase diagram
 shown in Fig. \ref{fig:depphase}, in which
the parameters deduced for
the spin-gap compounds $\rm CaV_4O_9$ and $\rm SrCu_2(BO_3)_2$
are also indicated. It is clearly seen how the 2D Heisenberg
model on the 1/5-depleted square lattice
is connected to the Shastry-Sutherland model.
It is particularly interesting to notice
 that a RVB-type spin gap state proposed
for the Shastry-Sutherland model is adiabatically connected to a
plaquette singlet state for $\rm CaV_4O_9$.

\subsection{Excitation spectrum}

We have seen that the spin-gap phases in the Shastry-Sutherland
model are continuously connected to the spin-gap phases
known for two different models. According to this relationship,
it may be expected that the  Shastry-Sutherland model shares some
of its  characteristic features with the latter models. As an example,
we here show that the spin-triplet excitation spectrum for the
RVB-type phase in the Shastry-Sutherland model
indeed exhibits a similar behavior expected for the
plaquette phase in the orthogonal-dimer chain.
\begin{figure}[htb]
\vspace{0.1cm}
\epsfxsize=7cm
\centerline{\epsfbox{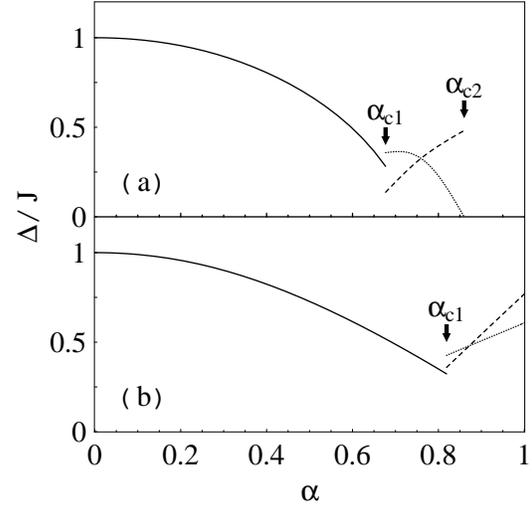}}
\caption{Spin-triplet excitation energy
$\Delta/J$ for (a) Shastry-Sutherland model and
(b) orthogonal-dimer chain as a function of $\alpha(=J'_{1}/J)$.
In (a), a triplet excitation spectrum \cite{Wei} over the dimer state
is plotted for $\alpha<\alpha_{c1}$, while two lowest triplet
excitations are shown for the RVB-type phase
$\alpha_{c1}<\alpha<\alpha_{c2}$.  In (b), similar plot is
done for the dimer phase ($\alpha<\alpha_c$) and also for
the plaquette phase ($\alpha_c <\alpha$).\cite{KogaB}
The meaning of the solid, dashed and dotted lines
is mentioned in the text.}
\label{fig:gap}
\end{figure}
In Fig. \ref{fig:gap}(a), we show the spin-triplet
excitation spectrum calculated for the 2D Shastry-Sutherland model.
In contrast to the dimer phase, where
the lowest triplet excitation is an almost-localized
triplet over the dimer ground state,\cite{Ueda1} we find that
the lowest-triplet excitation in the RVB-type phase
shows a more complicated feature; two triplet
excitations show the level-crossing around value of $\alpha' \simeq 0.76$.
It is not easy to clearly see the origin
of this level-crossing  because our calculation for the 2D model is
restricted to lower-order perturbation. Fortunately,
we can exploit the idea of the adiabatic continuation
by continuously changing  the system to the
orthogonal-dimer chain.

In the orthogonal-dimer chain, a similar level-crossing was
recently observed,\cite{KogaB} as shown in Fig. \ref{fig:gap}(b).
In this case, the nature of two different excitations in the
plaquette phase is well understood. Namely, in the lowest
excitation denoted by the dotted
line in the region of $\alpha>\alpha'(\simeq 0.87)$
is  a triplet excitation that simply breaks a plaquette singlet.
On the other hand  in the region $\alpha_{c1}<\alpha<\alpha'$,
the lowest-energy excited state shown by the dashed line
is described by a four-fold degenerate level
 without dispersion,\cite{KogaB}
in which two diagonal spins on a square
 plaquette becomes completely free,
giving rise to the four-fold degeneracy.\cite{KogaB}
The emergence of this peculiar excitation reflects the fact that the
system is located closely to the first-order phase transition point.
In other words, this level crossing is due to the strong
frustration effect around the transition point.
When the interchain coupling $k$ is introduced, the above two-types
of excitations are continuously
changed  to those in the Shastry-Sutherland model, as
shown in Fig. \ref{fig:gap}(a). In particular,
 four-fold degenerate states in the chain system
split into  singlet and triplet states, and the latter
gives the lowest-excited state shown by the dashed
line in the Shastry-Sutherland model in Fig. \ref{fig:gap}(a).
In this way, the characteristic level-crossing found in the
1D system persists even for the 2D system, from which
we can  say that the unusual level-crossing
found in the Shastry-Sutherland model reflects
the strong frustration  around the
first-order phase transition point.


\section{Summary}

We have studied the ground-state phase diagram
for a frustrated quantum spin model on the square lattice, which
systematically describes the systems for the
spin-gap compounds such as $\rm SrCu_2(BO_3)_2$ and $\rm CaV_4O_9$.
By calculating the ground state energy, the staggered
susceptibility and the spin gap by means of the series expansion method,
we have determined the phase diagram, and clarified
the nature of the associated quantum phase transitions.
One of the main purpose of the present study is to uncover
the origin of the intermediate
spin-gap state proposed for the Shastry-Sutherland model, which
has a spatially extended nature.
Starting from two types of the well-studied models, we have
observed how this intermediate phase emerges in the phase diagram.
Namely, with the use of the simple orthogonal-dimer chain model,
it has been  checked that the
first-order phase transition between the dimer and RVB-type phases
in the Shastry-Sutherland model has the same origin as in the transition
between the dimer and plaquette phases known for the chain system.
Further analysis based on the second model has allowed us to
show that the RVB-type intermediate state is adiabatically
connected to the plaquette state known for the 1/5-depleted square
lattice model, which describes the essential properties of
$\rm CaV_4O_9$. We have also shown that
the level-crossing of the triplet excitations in the Shastry-Sutherland model
is  understood  well in terms of their counterparts in the
1D orthogonal-dimer chain model.

These systematic investigations based on the adiabatic
continuation support that the intermediate
spin-gap phase for the Shastry-Sutherland model
 may be stabilized by geometrical frustration
induced by the competing interactions. However,
we wish to mention that there still remain unresolved
problems for the phase diagram of the Shastry-Sutherland model.
For example,
 Weihong {\it et al.} have shown by means of the Ising expansion\cite{Wei}
that the system has  the finite staggered magnetization down to
$(J'/J)\simeq 0.71$,
which may contradict the fact that the spin gap phase
is extended up to $(J'/J)\simeq 0.86 $.  Although we have checked
that the ground state energy of the
present spin-gap phase for $(J'/J) < 0.86 $
is indeed lower than that of Weihong
{\it et al.}, this apparent inconsistency
for the magnetization should be resolved in the future study.
Also, Knetter {\it et al.} have recently claimed that the instability of
the two-magnon excitation in the dimer phase may occur at $(J'/J)=0.630(5)$.
\cite{2trip} Although it may not be
clear whether this indeed  triggers a first-order  phase
transition to some new phase,\cite{2trip} it is an important
open problem to answer the above points for the phase diagram
of the Shastry-Sutherland
model. This is now under consideration.

The work is partly supported by a
Grant-in-Aid from the Ministry of Education, Science, Sports, and Culture.
A.K. is supported by the Japan Society for the Promotion of Science.



\begin{thebibliography}{99}

\bibitem{Kage}
H. Kageyama, K. Yoshimura, R. Stern, N. V. Mushnikov,
K. Onizuka, M. Kato, K. Kosuge, C. P. Slichter, T. Goto and
Y. Ueda,
Phys. Rev. Lett.  {\bf 82} (1999) 3168.

\bibitem{depE1}
S. Taniguchi, Y. Nishikawa, Y. Yasui, Y. Kobayashi,
M. Sato, T. Nishioka, M. Kontani and K. Sano,
J. Phys. Soc. Jpn. {\bf 64} (1995) 2758.

\bibitem{S-S}
B. S. Shastry and B. Sutherland,
Physica  {\bf 108B} (1981) 1069.

\bibitem{Ueda1}
S. Miyahara and K. Ueda,
Phys. Rev. Lett. {\bf 82} (1999) 3701.

\bibitem{Mila}
M. Albrecht and F. Mila,
Europhys. Lett. {\bf 34} (1996) 145.

\bibitem{KogaL}
A. Koga, and N. Kawakami,
Phys. Rev. Lett. {\bf 84} (2000) 4461.

\bibitem{2trip}
C. Knetter, A. B$\rm{\ddot{u}}$hler, E. M.-Hartmann
and G. S. Uhrig,
Phys. Rev. Lett. {\bf 85} (2000) 3958.



\bibitem{depT1}
K. Ueda, H. Kontani, M. Sigrist and P. A. Lee,
Phys. Rev. Lett. {\bf 76} (1996) 1932;
N. Katoh and M. Imada,
J. Phys. Soc. Jpn. {\bf 64} (1995) 4105.

\bibitem{Metap1}
Y. Fukumoto and A. Oguchi,
J. Phys. Soc. Jpn. {\bf 67} (1998) 2205.
\bibitem{Metap2}
Z. Weihong, J. Oitmma and C. J. Hamer,
Phys. Rev. B {\bf 58} (1998) 14147.

\bibitem{KogaB}
A. Koga, K. Okunishi and N. Kawakami,
Phys. Rev. B {\bf 62} (2000) 5558.

\bibitem{depT3}
M. Troyer, H. Kontani and K. Ueda,
Phys. Rev. Lett. {\bf 76} (1996) 3822.
\bibitem{Gel1}
R. R. P. Singh, M. P. Gelfand and D. A. Huse,
Phys. Rev. Lett. {\bf 61} (1988) 2484.
\bibitem{Wei}
Z. Weihong, C. J. Hamer and J. Oitmma,
Phys. Rev. B {\bf 60} (1999) 6608.
\bibitem{Pade}
A. J. Guttmann, in {\it{Phase Transitions and Critical Phenomena}},
edited by C. Domb and J. L. Lebowitz (Academic, New York, 1989)
Vol. 13.

\bibitem{3Dcl}
S. Chakravarty, B. I. Halperin and D. R. Nelson,
Phys. Rev. Lett. {\bf 60} (1988) 1057;
Phys. Rev. B {\bf 39} (1989) 2344.

\bibitem{Crit}
M. Ferer and A. Hamid-Aidinejad,
Phys. Rev. B {\bf 34} (1986) 6481.


\bibitem{Ivanov}
N. B. Ivanov and J. Richter,
Phys. Lett. {\bf 232A} (1997) 308;
J. Richter, N. B. Ivanov and J. Schulenburg,
J. Phys. Condence Matt. {\bf 10} (1998) 3635.

\bibitem{Ueda2}
S. Miyahara and K. Ueda,
J. Phys. Soc. Jpn. Suppl. B {\bf 69} (2000) 72.






\bibitem{Neelb}
Y. Fukumoto and A. Oguchi,
J. Phys. Soc. Jpn. {\bf 67} (1998) 697;
J. Phys. Soc. Jpn. {\bf 67} (1998) 2205;
Z. Weihong, J. Oitmaa and C. J. Hamer,
Phys. Rev. B {\bf 58} (1998) 14147;
R. R. P. Singh, Z. Weihong, C. J. Hamer and J. Oitmaa,
Phys. Rev. B {\bf 60} (1999) 7278;
A. Koga, S. Kumada and N. Kawakami,
J. Phys. Soc. Jpn. {\bf 68} (1999) 2373; {\bf 69} (2000) 1843.


\end{thebibliography}
\end{document}